\documentclass[fleqn,11pt]{wlscirep}
\usepackage[utf8]{inputenc}
\pdfoutput=1
\usepackage[T1]{fontenc}
\usepackage{algorithm}
\usepackage{algorithmic}
\usepackage{dirtytalk}

\usepackage{doi}
\usepackage{graphicx}
\usepackage{multicol}
\usepackage{multirow} 

\def\BibTeX{Bib\TeX}
\def\BibTeX{{\rm B\kern-.05em{\sc i\kern-.025em b}\kern-.08em
    T\kern-.1667em\lower.7ex\hbox{E}\kern-.125emX}}

\usepackage{xcolor}  
\usepackage{helvet}

\usepackage{cite}
\usepackage{amsmath,amssymb,amsfonts,color,colortbl,textgreek}

\usepackage{lettrine,paralist,fancyhdr}
\usepackage{nomencl,makeidx,pdfpages}
\usepackage{setspace}

\usepackage{longtable}
\usepackage{booktabs} 

\usepackage{bigdelim} 
\usepackage{bigstrut} 
\usepackage{tabularx}
\usepackage{ragged2e}
 \usepackage{float}

\usepackage{textcase}
\usepackage{listings}

\usepackage{pgfplots}
\usepackage{pgfplotstable}
\usepackage{tikz}
\pgfplotsset{compat=newest}
\pgfplotsset{compat=1.18}
\usepackage{eqparbox}

\usepackage{siunitx}
 \usepackage{fancyhdr}
\DeclareSymbolFont{NewLetters}{OML}{ntxmi}{m}{it}

\usepackage[labelfont=bf]{caption}
\usepackage{subcaption}
\usepackage{varioref}

\usepackage{geometry}
\usepackage{sansmathfonts}

\tikzstyle{process} = [rectangle, rounded corners, minimum width=3cm, minimum height=1cm, text centered, draw=black, fill=blue!20]
\tikzstyle{arrow} = [thick,->,>=stealth]

\tikzstyle{process} = [rectangle, rounded corners, minimum width=2.5cm, minimum height=1cm, text centered, draw=black, fill=blue!20, text width=2.5cm]
\tikzstyle{arrow} = [thick,->,>=stealth]

\tikzstyle{process} = [rectangle, rounded corners, minimum width=2.5cm, minimum height=1cm, text centered, draw=black, fill=blue!20, text width=2.5cm]
\tikzstyle{arrow} = [thick,->,>=stealth]

\def\BibTeX{{\rm B\kern-.05em{\sc i\kern-.025em b}\kern-.08em
    T\kern-.1667em\lower.7ex\hbox{E}\kern-.125emX}}

\usepackage[labelfont=bf]{caption}
\usepackage{subcaption}
\usepackage{varioref}

\usepackage[numbers]{natbib} 
\setlist[itemize]{itemsep=-5pt, topsep=0pt} 

\usepackage{hyperref} 
\usepackage{orcidlink} 

\usepackage{textcomp}
\usepackage{bm}
\makeatletter
\AtBeginDocument{\DeclareMathVersion{bold}
\SetSymbolFont{operators}{bold}{T1}{times}{b}{n}
\SetSymbolFont{NewLetters}{bold}{T1}{times}{b}{it}
\SetMathAlphabet{\mathrm}{bold}{T1}{times}{b}{n}
\SetMathAlphabet{\mathit}{bold}{T1}{times}{b}{it}
\SetMathAlphabet{\mathbf}{bold}{T1}{times}{b}{n}
\SetMathAlphabet{\mathtt}{bold}{OT1}{pcr}{b}{n}
\SetSymbolFont{symbols}{bold}{OMS}{cmsy}{b}{n}
\renewcommand\boldmath{\@nomath\boldmath\mathversion{bold}}}
\makeatother

\def\BibTeX{{\rm B\kern-.05em{\sc i\kern-.025em b}\kern-.08em
    T\kern-.1667em\lower.7ex\hbox{E}\kern-.125emX}}

\title{The Future of IPTV: Security, AI Integration, 5G, and Next-Gen Streaming}

\title{The Future of IPTV: Security, AI Integration, 5G, and Next-Gen Streaming}
\author[1]{Georgios Giannakopoulos\orcidlink{0000-0002-3707-3276}}
\author[2]{Peter Adegbenro\orcidlink{0009-0000-2524-873X}}
\author[3]{Maria Antonnette Perez\orcidlink{0009-0007-8436-4774}}

\affil[1]{Independent Researcher, The Hague, The Netherlands} 
\affil[2]{Independent Researcher, Ilorin, Nigeria} 
\affil[3]{Independent Researcher, Manila, Philippines}

\begin{abstract}
The evolution of Internet Protocol Television (IPTV) has transformed the landscape of digital broadcasting by leveraging high-speed internet connectivity to deliver high-quality multimedia content. IPTV provides a dynamic and interactive television experience through managed networks, ensuring superior Quality of Service (QoS) compared to open-network Internet TV. This study explores the technical infrastructure of IPTV, including its network architecture, data compression techniques, and the role of protocols such as IGMP and RTSP. It also examines security challenges, including encryption, digital rights management (DRM), and authentication mechanisms that safeguard IPTV services from unauthorized access and piracy.

Moreover, the paper analyzes the distinctions between IPTV and open-network Internet TV, highlighting their respective advantages and limitations in terms of service control, bandwidth optimization, and content security. The integration of artificial intelligence (AI) and machine learning (ML) in IPTV enhances personalized content recommendations and predictive analytics, leading to improved user engagement and efficient network management. Additionally, emerging technologies such as 5G and cloud-based IPTV services are explored for their potential to further revolutionize the industry.

While IPTV presents a robust alternative to traditional broadcasting, challenges such as bandwidth constraints, cybersecurity threats, and regulatory compliance remain significant. The study concludes that IPTV’s future success will depend on advancements in network infrastructure, AI-driven optimizations, and strategic regulatory adaptations. As IPTV continues to evolve, hybrid models integrating IPTV and open-network streaming services are expected to enhance content accessibility, security, and overall user experience.
\end{abstract}
\begin{document}

\flushbottom
\maketitle
%
%
\thispagestyle{empty}


\section{Introduction}
\label{section1}
The advent of high-speed internet and advancements in multimedia streaming technologies have fueled the growth of IPTV \cite{Prodani2024}. Unlike traditional broadcasting methods, IPTV relies on internet protocols to deliver content, allowing for greater flexibility, interactivity, and service bundling, such as triple play (voice, internet, and TV) and quadruple play (adding mobile services) \cite{AlMajeed2024}. However, its adoption remains hindered by several technical, infrastructural, and regulatory challenges \cite{Giannakopoulos202503}, \cite{Mikulec2019}.

One of the defining characteristics of IPTV is its ability to provide on-demand services, allowing users to access a diverse range of content at their convenience \cite{Thakore2024}. Unlike traditional broadcast television, where content follows a predefined schedule, IPTV allows users to watch content whenever they choose, including live broadcasts, time-shifted television, and video-on-demand (VoD) services \cite{Khan2023}.

Additionally, IPTV operates on a managed network, meaning that service providers control the delivery process, ensuring a higher quality of service (QoS) compared to open-network streaming services \cite{IMARC2024}. This controlled environment enables better security, content protection, and reliable streaming, reducing buffering and latency issues common with open-network Internet TV services \cite{Almohamad2021}.

Despite its benefits, IPTV faces substantial challenges. Bandwidth limitations \cite{deAzambuja2023}, network congestion \cite{ITU2023}, interoperability concerns \cite{Lee2019}, and regulatory compliance issues remain barriers to widespread adoption \cite{Basso2011}. Furthermore, security vulnerabilities such as unauthorized access \cite{Riskoria2024}, data privacy concerns \cite{Geng2007}, and digital piracy \cite{Katsikas2024} pose risks to both service providers and consumers. This paper aims to explore these challenges while also discussing advancements that can enhance IPTV services, making them more efficient, secure, and widely adopted \cite{Zhang2022}.

\section{Technical Framework of IPTV}
\label{section2}
\subsection{IPTV Network Architecture}
\label{subsection2.1}
\begin{figure}[h]
\begin{center}
{
\includegraphics[width=0.6\textwidth]{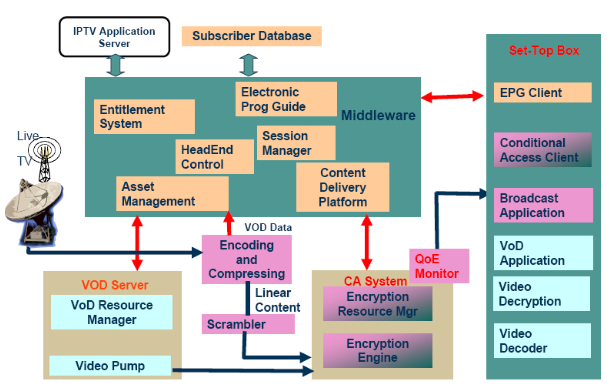}
}
         \centering
         \caption{Simplified IPTV system \cite{Punchihewa2011}, \cite{Benoit_2007}.}
         \label{fig1}
      \end{center}   
\end{figure}

IPTV operates on a client-server model, where video content is transmitted over managed networks, ensuring controlled Quality of Service (QoS) \cite{Kumar2021}. The infrastructure consists of content providers, head-end servers, middleware platforms, and end-user devices \cite{Chen2020}. Head-end servers encode and distribute content, middleware handles authentication and service delivery, and set-top boxes (STBs) or smart TVs act as the client interface \cite{Patel2019}. The network topology includes multicasting, caching mechanisms, and adaptive streaming to optimize performance and minimize latency \cite{Nguyen2018}. A Simplified IPTV system, is shown in \hyperref[fig1]{Figure 1} below \cite{Punchihewa2011}, \cite{Benoit_2007}.


\begin{figure}[ht!]
\begin{center}
{
\includegraphics[width=0.5\textwidth]{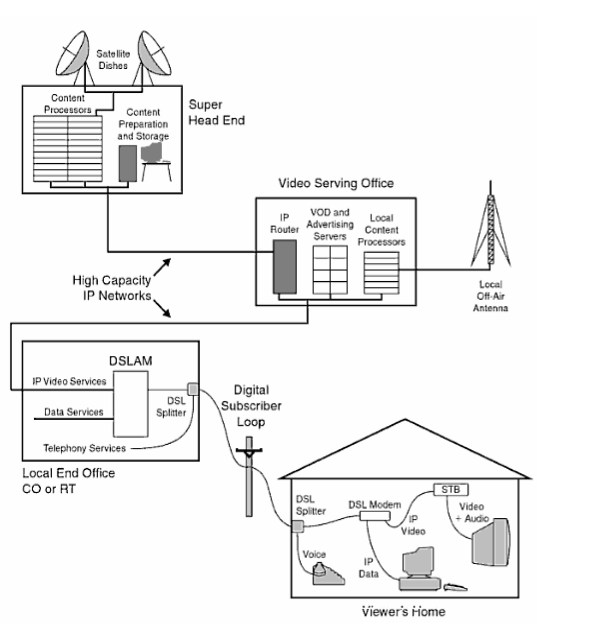}
}
         \centering
         \caption{IPTV basic structure \cite{Punchihewa2011}, \cite{Telecom_IPTV_2010}.}
         \label{fig2}
      \end{center}   
\end{figure}

A typical IPTV architecture is divided into three main layers: content acquisition, distribution, and end-user access \cite{Lee2017}. The content acquisition layer involves capturing, encoding, and compressing video streams from various sources \cite{Gonzalez2016}. The distribution layer transmits these compressed streams over a managed IP network, ensuring efficient data flow and minimal latency \cite{Johnson2015}. The end-user access layer consists of the consumer’s viewing device, whether a smart TV, set-top box, computer, or mobile device, which decodes and presents the content for viewing \cite{Anderson2014}. An IPTV basic structure, is shown in \hyperref[fig2]{Figure 2} below \cite{Punchihewa2011}, \cite{Telecom_IPTV_2010}.

Moreover, IPTV architecture supports both unicast and multicast transmission modes \cite{Huang2013}. Unicast transmission delivers content to individual users, making it suitable for VoD services, while multicast transmission efficiently distributes live television broadcasts to multiple users simultaneously \cite{Wilson2022}. The implementation of efficient network management techniques, such as content delivery networks (CDNs) \cite{Santos2021} and edge computing \cite{Smith2023}, further enhances IPTV scalability and performance.

\subsection{Data Compression and Encoding Techniques}
\label{subsection2.2}

To reduce bandwidth consumption while maintaining high video quality, IPTV employs compression algorithms such as MPEG-2, H.264 (AVC), and the more recent H.265 (HEVC) \cite{Aspen2024}. These codecs use predictive and spatial-temporal redundancy techniques to decrease data rates without significantly compromising visual fidelity \cite{Fortinet2023}. The transition from MPEG-2 to H.265 has led to a \(50\%\) reduction in required bandwidth while maintaining similar video quality \cite{ZhangX2014}.     

Data compression plays a crucial role in IPTV by optimizing network efficiency and minimizing storage requirements \cite{Giannakopoulos202503}. The use of inter-frame and intra-frame compression techniques ensures that only essential video information is transmitted, reducing redundancy \cite{Prodani2024}, \cite{Chen2020}. While MPEG-2 remains widely used for standard-definition broadcasts, H.264 and H.265 have become the preferred choices for high-definition (HD) and ultra-high-definition (UHD) content due to their superior compression capabilities \cite{Giannakopoulos202503}, \cite{Johnson2015}.

Furthermore, adaptive bitrate streaming (ABS) techniques allow IPTV systems to dynamically adjust video quality based on real-time network conditions \cite{Huang2013}. This ensures a seamless viewing experience by reducing buffering and adapting video resolution based on available bandwidth  \cite{Chen2020}. Innovations such as AV1 and VVC (Versatile Video Coding) are expected to further improve compression efficiency, enabling IPTV providers to deliver 4K and 8K content more effectively \cite{Patel2019}.

\subsection{Internet Group Management Protocol (IGMP) and Real-Time Streaming Protocol (RTSP)}
\label{subsection2.3}

IPTV services employ IGMP for multicast streaming, allowing multiple users to watch the same live broadcast without excessive bandwidth consumption \cite{Patel2019}, \cite{Mondal518629}. RTSP, on the other hand, enables video-on-demand (VoD) features such as pause, rewind, and fast forward \cite{Nguyen2018}, \cite{Momeni2015}. These protocols optimize data delivery by reducing network congestion and enhancing the viewer experience \cite{Gonzalez2016}.

IGMP enables efficient multicast communication by allowing devices to join and leave multicast groups dynamically \cite{Huang2013}. This reduces bandwidth consumption by sending a single video stream to multiple users instead of duplicating transmissions for each recipient \cite{Geng2007}. IGMP versions 1, 2, and 3 have introduced improvements in channel switching speeds and network efficiency \cite{Prodani2024}, \cite{AlMajeed2024}, \cite{ZhangX2014}.

RTSP facilitates VoD services by enabling direct control over streaming sessions \cite{Mondal518629}. Users can manipulate playback, seek content, and pause live streams using RTSP commands \cite{Nguyen2018}. The combination of IGMP and RTSP allows IPTV to support both live and on-demand content delivery efficiently \cite{Patel2019}, \cite{DVB_MHP_2000}. Future enhancements to streaming protocols, such as QUIC-based transmission and HTTP adaptive streaming, will further improve network reliability and user experience \cite{Lee2019}, \cite{Gonzalez2016}.

\subsection{Advanced Data Compression and Encoding Techniques}
\label{subsection2.4}
Advanced codecs play a crucial role in IPTV efficiency \cite{AlMajeed2024}, \cite{Giannakopoulos202503}. H.264 (AVC) provides efficient compression and is widely adopted, while H.265 (HEVC) offers even better compression, making high-resolution video streaming feasible over existing broadband connections \cite{Zhang2022}, \cite{Punchihewa2011}. Future codecs like AV1 and VVC promise further efficiency gains, enabling 4K and 8K streaming with reduced network overhead \cite{Chen2020}, \cite{DVB_MHP_2000}.

H.264 revolutionized video streaming by reducing bandwidth consumption without significantly compromising video quality \cite{Patel2019}. However, with the growing demand for UHD content, H.265 (HEVC) emerged as a superior alternative, offering double the compression efficiency of its predecessor \cite{Lee2017}. The adoption of HEVC has allowed IPTV services to deliver 4K and HDR (High Dynamic Range) content with minimal bandwidth requirements \cite{ZhangX2014}.

The introduction of AV1, an open-source codec developed by the Alliance for Open Media (AOM), has further disrupted the industry by offering better compression rates than HEVC without requiring licensing fees \cite{Riskoria2024}. Additionally, VVC, also known as H.266, aims to provide a \(50\%\) bitrate reduction compared to HEVC while maintaining equivalent quality \cite{Zhang2022}. These advancements in codec technology are essential for supporting the next generation of IPTV services, reducing storage costs, and enhancing overall streaming efficiency \cite{Benoit_2007}.

\section{IPTV and Quality of Service (QoS)}
\label{section3}
\subsection{Bandwidth Requirements and Optimization}
\label{subsection3.1}

\begin{figure}[h]
\begin{center}
{
\includegraphics[width=0.8\textwidth]{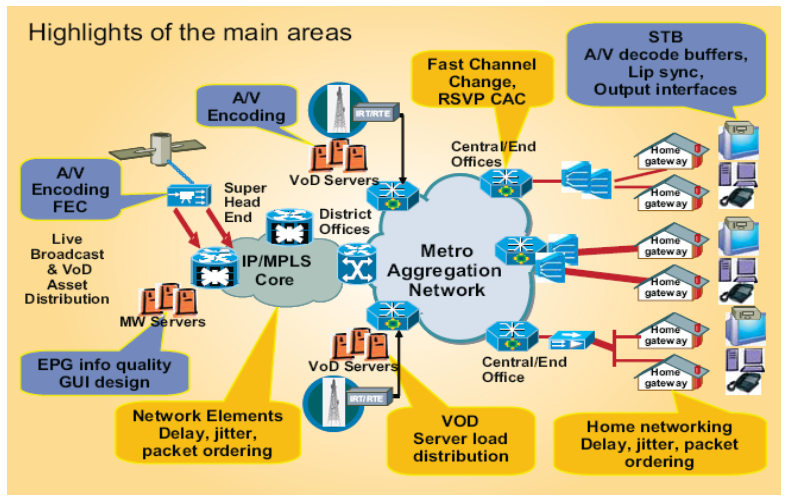}
}
         \centering
         \caption{IPTV QoE in the end-to-end model \cite{Punchihewa2011}, \cite{simpson_2005}.}
         \label{fig3}
      \end{center}   
\end{figure}

One of the primary concerns in IPTV deployment is ensuring adequate bandwidth to support HD and UHD video streaming \cite{Mikulec2019}. IPTV services demand a considerable amount of network bandwidth, which increases with video resolution, frame rate, and audio quality \cite{ITU2023}. For example, a standard-definition (SD) IPTV stream may require 2-3 Mbps, while a high-definition (HD) stream requires around 5-10 Mbps \cite{Patel2019}. Ultra-high-definition (UHD or 4K) streaming may need upwards of 25 Mbps per stream, necessitating high-speed broadband connections \cite{Chen2020}. An IPTV QoE in the end-to-end model, is shown in \hyperref[fig3]{Figure 3} below \cite{Punchihewa2011}, \cite{simpson_2005}.

To optimize bandwidth utilization, IPTV providers implement various compression techniques and network management strategies \cite{Prodani2024}. Compression standards like H.264 (AVC) and H.265 (HEVC) significantly reduce the data required for video transmission while preserving quality \cite{AlMajeed2024}, \cite{Giannakopoulos202503}. Additionally, the use of adaptive bitrate streaming (ABS) allows IPTV services to adjust video quality dynamically based on available bandwidth, preventing excessive buffering and ensuring smooth playback  \cite{ITU2023}, \cite{simpson_2005}.

Another approach to bandwidth optimization is the implementation of content delivery networks (CDNs) \cite{Lee2019}, \cite{Patel2019}. CDNs reduce network congestion by caching IPTV content in geographically distributed servers, delivering data from the closest possible location to the end user \cite{Patel2019}. Multicasting is also employed to efficiently deliver live broadcasts to multiple users without consuming excessive network resources \cite{Santos2021}. These strategies collectively enhance the scalability and efficiency of IPTV networks, ensuring seamless viewing experiences for users \cite{Gonzalez2016}, \cite{Mondal518629}.

\newpage 
\subsection{Latency and Packet Loss Mitigation}
\label{subsection3.2}

Latency and packet loss are critical concerns in IPTV service quality, particularly for live streaming applications \cite{Santos2021}. Latency refers to the time delay between content transmission and reception, and excessive latency can degrade user experience, especially for live sports broadcasts and interactive IPTV applications \cite{Telecom_IPTV_2010}, \cite{ZhangX2014}, \cite{simpson_2005}. Packet loss occurs when data packets are dropped due to network congestion, resulting in visual artifacts, freezing, or buffering issues \cite{Patel2019}, \cite{Benoit_2007}, \cite{Punchihewa2011}.

Several techniques help mitigate these problems. Forward error correction (FEC) is commonly used to recover lost packets by adding redundant data to the stream, allowing for reconstruction in the event of transmission errors  \cite{Alzoubi2016}. Real-time transport protocol (RTP) is another key technology used in IPTV to ensure smooth video transmission by detecting packet loss and requesting retransmissions when necessary \cite{Momeni2015},  \cite{claro_pereira_campos}.

Edge caching and adaptive streaming further contribute to minimizing latency by storing frequently accessed content closer to end users and adjusting playback quality in real time based on network conditions \cite{Abdollahpouri2012}. The implementation of low-latency protocols such as HTTP/2 and QUIC also helps in reducing transmission delays, enhancing the overall responsiveness of IPTV services \cite{Benoit_2007}.

\subsection{Adaptive Bitrate Streaming}
\label{subsection3.3}

Adaptive bitrate streaming (ABS) is a crucial technology in IPTV that dynamically adjusts video quality based on the user’s internet speed and network conditions \cite{Nguyen2018}. Unlike traditional streaming methods that deliver a fixed bitrate stream, ABS continuously monitors bandwidth availability and device capabilities, selecting the optimal bitrate for seamless playback \cite{Lee2017}.

ABS operates by encoding the same video content at multiple quality levels and storing them in different resolutions and bitrates \cite{Wilson2022}. When a user starts streaming, the ABS algorithm determines the most suitable quality level based on real-time network conditions \cite{Santos2021}. If bandwidth availability drops, the stream switches to a lower resolution to prevent buffering, and when bandwidth improves, the system upgrades the quality accordingly \cite{Anderson2014}.

The advantages of ABS include enhanced viewer experience, efficient bandwidth utilization, and greater compatibility across different devices and network conditions \cite{Aspen2024}. Modern IPTV platforms employ ABS alongside CDNs to distribute video content effectively, ensuring uninterrupted playback for users even under fluctuating network conditions \cite{Punchihewa2011}. Future advancements in ABS technology, including AI-driven predictive adjustments, promise further improvements in video streaming efficiency \cite{ZhangX2014}.

\section{IPTV Security Challenges and Solutions}
\label{section4}
\subsection{Encryption and Digital Rights Management (DRM)}
\label{subsection4.1}

Security is a fundamental concern in IPTV services, as content providers must protect intellectual property and prevent unauthorized access \cite{Thakore2024}. Encryption techniques such as AES-128 and AES-256 are widely used to secure IPTV transmissions, ensuring that only authorized users can decrypt and view the content \cite{Almohamad2021}.

Digital Rights Management (DRM) systems play a key role in IPTV security by controlling content access and distribution \cite{deAzambuja2023}. DRM solutions such as Microsoft PlayReady, Google Widevine, and Apple FairPlay employ encryption, user authentication, and license management to prevent piracy and unauthorized redistribution \cite{Riskoria2024}. These technologies ensure compliance with copyright regulations while safeguarding revenue for IPTV providers \cite{Kumar2021}.

Despite these measures, IPTV services remain vulnerable to security threats such as credential sharing, illegal restreaming, and hacking attempts \cite{Thakore2024}, \cite{deAzambuja2023}. Service providers continually enhance security protocols by integrating blockchain technology, watermarking techniques, and AI-driven fraud detection to identify and prevent illicit activities \cite{Almohamad2021}.

\subsection{User Authentication and Access Control}
\label{subsection4.2}
User authentication and access control mechanisms are essential for securing IPTV services and ensuring that only legitimate subscribers can access content \cite{Kumar2021}. Multi-factor authentication (MFA) is widely implemented, requiring users to verify their identity through multiple authentication steps, such as passwords, biometric recognition, or one-time passcodes \cite{ITU2023}.

Access control mechanisms such as geo-blocking restrict content availability based on a user's location, preventing unauthorized access from outside designated regions \cite{Patel2019}. IP whitelisting and blacklisting further enhance security by allowing or denying access based on predefined network policies \cite{ITU2023}.

Subscription-based IPTV services often implement token-based authentication, where temporary access tokens are issued to verified users, minimizing the risk of credential theft \cite{Riskoria2024}. By combining these authentication measures, IPTV providers strengthen platform security and ensure compliance with licensing agreements \cite{Riskoria2024}.

\subsection{Network Threats and Cybersecurity Strategies}
\label{subsection4.3}
IPTV platforms are susceptible to various cyber threats, including distributed denial-of-service (DDoS) attacks, malware infections, and unauthorized content distribution \cite{ITU2023}. DDoS attacks overwhelm IPTV servers with excessive traffic, causing service disruptions, while malware can compromise user data and network integrity \cite{Aspen2024}.

To counter these threats, IPTV providers deploy advanced cybersecurity measures such as firewalls, intrusion detection systems (IDS), and artificial intelligence (AI)-based threat monitoring \cite{Thakore2024}, \cite{deAzambuja2023}. AI-driven security tools analyze network traffic patterns in real-time, identifying anomalies and mitigating potential cyber threats before they cause harm \cite{vasanthi_chidambaram_2014}.

Additionally, encryption techniques, VPN-based content delivery, and secure API communication protect IPTV infrastructures from cyber-attacks. As cybersecurity threats evolve, IPTV providers must continually update security policies and invest in robust protective measures to safeguard user data and content integrity \cite{Aspen2024}, \cite{vasanthi_chidambaram_2014}.

\section{IPTV vs. Open-Network Internet TV}
\label{section5}
\subsection{Comparative Analysis of Transmission Methods}
\label{subsection5.1}

IPTV and open-network Internet TV are both digital television delivery methods, but they differ significantly in terms of transmission, quality, security, and infrastructure requirements \cite{Telecom_IPTV_2010}, \cite{DVB_MHP_2000}. IPTV operates over managed networks, ensuring controlled service delivery, while Internet TV relies on open, public internet connections, leading to potential quality and reliability issues \cite{Prodani2024}, \cite{Mondal518629}.

IPTV utilizes dedicated broadband infrastructure with multicast technology, allowing for optimized bandwidth usage and consistent video quality \cite{Mondal518629}. It supports features such as video-on-demand (VoD), interactive services, and high-definition (HD) and ultra-high-definition (UHD) streaming \cite{Benoit_2007}, \cite{DVB_MHP_2000}. On the other hand, Internet TV, including platforms like YouTube, Netflix, and Hulu, uses unicast transmission, where each viewer receives an individual video stream, often leading to increased bandwidth consumption and potential buffering \cite{Giannakopoulos202503}, \cite{Gonzalez2016}.

Another significant distinction is service control. IPTV providers have end-to-end control over content distribution, ensuring reliable Quality of Service (QoS) through network prioritization and error correction techniques \cite{Giannakopoulos202503}, \cite{Mikulec2019}. In contrast, Internet TV services rely on content delivery networks (CDNs) and adaptive bitrate streaming (ABS) to mitigate latency and buffering issues, but they lack the same level of control as IPTV  \cite{Nguyen2018}, \cite{Gonzalez2016}.

Security is also a critical differentiator. IPTV services implement strong encryption, digital rights management (DRM), and access control measures, whereas Internet TV platforms are more susceptible to piracy and unauthorized content distribution \cite{Lee2017}, \cite{Liu2007}. The closed-system nature of IPTV makes it a preferred choice for premium broadcasting, whereas open-network Internet TV is better suited for freely accessible or ad-supported content \cite{Gonzalez2016}.

\subsection{Closed vs. Open Networks: Pros and Cons}
\label{subsection5.2}

The choice between closed IPTV networks and open Internet TV networks depends on multiple factors, including service reliability, cost, accessibility, and content security \cite{Huang2013}.

Closed IPTV networks provide a highly controlled environment where service providers can guarantee QoS, minimal latency, and secure content distribution \cite{Patel2019}. Because IPTV operates on managed networks, it is less prone to congestion and buffering, making it ideal for delivering premium television services, live sports, and VoD \cite{Benoit_2007}, \cite{DVB_MHP_2000}. However, this control comes with increased costs for infrastructure deployment and maintenance, making it less accessible for independent content creators \cite{Lee2019}.

Open-network Internet TV, on the other hand, offers greater flexibility and lower barriers to entry, allowing anyone with an internet connection to stream or broadcast content \cite{AlMajeed2024}. It supports user-generated content, enabling individuals and businesses to distribute video without requiring proprietary infrastructure \cite{Basso2011}. However, open networks come with challenges such as inconsistent streaming quality, higher vulnerability to cyber threats, and weaker content protection mechanisms \cite{deAzambuja2023}.

Ultimately, IPTV’s closed-system approach is advantageous for broadcasters seeking stability and security, whereas open-network Internet TV provides broader accessibility and affordability for general users and independent creators \cite{Chen2020}, \cite{Patel2019}. The evolution of hybrid models integrating both approaches could offer a balanced solution that maximizes service reliability while maintaining accessibility \cite{Huang2013}, \cite{Santos2021}.

\section{The Role of AI and ML in IPTV}
\label{section6}
\subsection{Personalized Content Recommendation Systems}
\label{subsection6.1}
Artificial intelligence (AI) \cite{Thakore2024}, \cite{deAzambuja2023}, \cite{Wilson2022}, \cite{Aspen2024}, \cite{Chaitanya2023} and machine learning (ML) \cite{Zhang2022}, \cite{Chaitanya2023} are transforming IPTV by enhancing user experiences through personalized content recommendations \cite{Geng2007}, \cite{Katsikas2024}. By analyzing viewing habits, preferences, and behavioral patterns, AI-powered recommendation engines suggest content tailored to individual users, increasing engagement and retention rates \cite{Lee2019}, \cite{Mondal518629}.

Machine learning models process vast amounts of data, including watch history, search queries, and demographic information, to predict what content users are likely to enjoy \cite{Geng2007}, \cite{Gonzalez2016}. These models employ collaborative filtering, content-based filtering, and deep learning algorithms to refine recommendations \cite{ITU2023}. Platforms like Netflix and Amazon Prime Video use these techniques to curate personalized playlists and improve user satisfaction \cite{Giannakopoulos202503}.

AI-driven recommendation systems also benefit IPTV service providers by optimizing content distribution strategies \cite{Mikulec2019}. By identifying trending content and peak viewing times, providers can allocate network resources more efficiently and ensure seamless content delivery \cite{Geng2007}, \cite{Patel2019}. Additionally, AI enhances advertising strategies by targeting users with relevant ads based on their preferences, improving ad revenue generation \cite{IMARC2024}, \cite{Zhang2022}.

Future advancements in AI are expected to further enhance IPTV personalization, incorporating real-time feedback mechanisms and sentiment analysis to refine recommendations dynamically \cite{Zhang2022}, \cite{Chen2020}. These innovations will continue to drive user engagement and retention in IPTV services \cite{Chaitanya2023}, \cite{Haque2023}.

\subsection{Predictive Analytics for Network Optimization}
\label{subsection6.2}
Predictive analytics, powered by AI and ML, plays a crucial role in optimizing IPTV network performance by analyzing historical data and forecasting network traffic patterns \cite{Mondal518629}. By anticipating congestion, bandwidth requirements, and potential service disruptions, IPTV providers can take proactive measures to enhance QoS \cite{Zhang2022}, \cite{Huang2013}.

One key application of predictive analytics is traffic load balancing \cite{Zhang2022}. AI algorithms monitor network usage in real time and dynamically adjust data routing to minimize latency and prevent bottlenecks \cite{Kumar2021}, \cite{Patel2019}. This ensures a consistent and smooth viewing experience, particularly during peak usage hours \cite{Nguyen2018}, \cite{Gonzalez2016}.

Predictive maintenance is another valuable application, where AI-driven systems analyze equipment performance data to detect potential failures before they occur \cite{Thakore2024}, \cite{deAzambuja2023}, \cite{Wilson2022}, \cite{simpson_2005}. This proactive approach minimizes downtime, reduces maintenance costs, and improves service reliability \cite{Riskoria2024}, \cite{DVB_MHP_2000}.

Additionally, AI enhances content caching strategies by predicting popular content trends and preloading frequently watched videos closer to end users \cite{Lee2019}, \cite{Mondal518629}. This reduces the load on central servers and enhances streaming speeds, particularly in geographically distributed IPTV networks \cite{Patel2019}, \cite{Mondal518629}.

As AI technologies continue to evolve, predictive analytics will become an integral part of IPTV service management, ensuring optimal performance, cost efficiency, and a superior user experience \cite{Geng2007}, \cite{ZhangX2014}.

\section{Emerging Trends and Future Directions}
\label{section7}
\subsection{Integration with 5G Networks}
\label{subsection7.1}

The integration of 5G technology is set to revolutionize IPTV by providing higher bandwidth, lower latency, and greater network efficiency \cite{Haque2023}. With speeds reaching up to 10 Gbps and ultra-low latency, 5G will enable seamless UHD and 8K streaming, eliminating buffering and lag issues \cite{Khan2023}, \cite{Smith2023}.

One major advantage of 5G for IPTV is its ability to support massive device connectivity \cite{Zhang2022}, \cite{Smith2023}. This is particularly beneficial for multi-device streaming environments, where households have multiple smart TVs, mobile devices, and IoT-connected systems accessing IPTV services simultaneously \cite{Khan2023}, \cite{Haque2023}.

Additionally, network slicing, a key feature of 5G, allows service providers to allocate dedicated bandwidth for IPTV applications, ensuring high-priority transmission without interference from other internet traffic \cite{Almohamad2021}, \cite{Haque2023}. This guarantees uninterrupted streaming, even in congested network conditions \cite{Smith2023}.

The adoption of 5G-enabled IPTV will also enhance mobile streaming experiences, making it easier for users to access high-quality content on the go  \cite{Chen2020}, \cite{Patel2019}. With the widespread deployment of 5G, IPTV services will become more efficient, scalable, and immersive \cite{Huang2013}, \cite{Smith2023}, \cite{Haque2023}.

\subsection{Cloud-Based IPTV Services}
\label{subsection7.2}

Cloud computing is playing an increasingly important role in IPTV, enabling service providers to deliver scalable, flexible, and cost-effective streaming solutions \cite{Wilson2022}, \cite{Haque2023}. Cloud-based IPTV eliminates the need for expensive on-premise infrastructure by leveraging remote data centers for content storage, processing, and distribution \cite{Santos2021}, \cite{claro_pereira_campos}.

One of the key benefits of cloud-based IPTV is its ability to scale dynamically based on demand \cite{Mondal518629}, \cite{Momeni2015}. Service providers can instantly expand server capacity during peak hours and reduce it during low-traffic periods, optimizing resource utilization and reducing operational costs \cite{Riskoria2024}, \cite{claro_pereira_campos}.

Additionally, cloud-based solutions facilitate content delivery via CDNs, reducing latency and enhancing global accessibility \cite{Momeni2015}, \cite{vasanthi_chidambaram_2014}. AI-driven analytics in the cloud also help optimize content recommendations, predictive maintenance, and security measures \cite{Riskoria2024}, \cite{Katsikas2024}, \cite{Aspen2024}.

With the rise of edge computing, IPTV services are further enhancing performance by processing data closer to end users \cite{Lee2019}, \cite{Geng2007}. This reduces network congestion and improves streaming speeds, ensuring a superior viewing experience \cite{Gonzalez2016}, \cite{Mondal518629}.

\subsection{Regulatory and Market Trends}
\label{subsection7.3}
The IPTV industry is subject to evolving regulatory frameworks and market dynamics \cite{IMARC2024}. Governments and regulatory bodies worldwide are establishing policies to address content licensing, data privacy, and net neutrality, shaping the future of IPTV services \cite{ITU2023}.

One key regulatory challenge is ensuring fair competition among IPTV providers while preventing monopolistic control over content distribution \cite{Johnson2015}, \cite{Wilson2022}. Additionally, policies surrounding digital rights management (DRM) and anti-piracy measures are critical to protecting content creators and broadcasters \cite{Lee2017}, \cite{Liu2007}.

Market trends indicate a growing preference for subscription-based IPTV models, with increasing adoption of hybrid monetization strategies that combine advertisements and paid subscriptions \cite{Wilson2022}. As user demand for high-quality, personalized content grows, IPTV providers must continually adapt to regulatory changes and market expectations \cite{IMARC2024}, \cite{Riskoria2024}.

\section{Conclusion-Discussion}
\label{section8}
The IPTV industry has experienced rapid evolution, driven by advancements in internet infrastructure, data compression technologies, and AI-driven service enhancements \cite{Mikulec2019}. IPTV provides a flexible and scalable alternative to traditional broadcasting, offering personalized content delivery, enhanced security, and interactive experiences for consumers \cite{Thakore2024}, \cite{deAzambuja2023}. However, several technical, operational, and regulatory challenges remain \cite{Giannakopoulos202503}, \cite{Chen2020}.

One of the key challenges in IPTV adoption is bandwidth optimization \cite{Patel2019}, \cite{Nguyen2018}. While compression technologies like H.265 (HEVC) and emerging codecs such as AV1 have significantly reduced data consumption, the increasing demand for UHD and 8K content continues to put pressure on network infrastructure \cite{Zhang2022}. The integration of 5G and cloud-based solutions offers promising avenues for scalability and enhanced content delivery \cite{Zhang2022}, \cite{Patel2019}.

Security remains a critical concern, with IPTV providers constantly improving encryption protocols, digital rights management (DRM), and authentication mechanisms to combat piracy and unauthorized content access \cite{ITU2023}, \cite{Lee2017}. The adoption of AI in predictive security analytics further strengthens protection against cyber threats and unauthorized data access \cite{Kumar2021}, \cite{Johnson2015}, \cite{Anderson2014}.

Regulatory frameworks will also shape the future of IPTV, as governments and industry organizations impose content licensing, net neutrality, and anti-piracy measures \cite{Thakore2024}, \cite{ITU2023}, \cite{Santos2021}, \cite{vasanthi_chidambaram_2014}. These regulations will impact market competitiveness and revenue generation models, requiring service providers to adapt to evolving compliance requirements \cite{Thakore2024}, \cite{Almohamad2021}, \cite{deAzambuja2023}.

Despite these challenges, IPTV is poised to become a dominant force in digital entertainment \cite{Lee2019}, \cite{Geng2007}. The increasing reliance on AI-driven personalization, predictive analytics, and cloud-based infrastructures will continue to enhance IPTV’s efficiency and user experience \cite{Gonzalez2016}, \cite{Huang2013}. Moreover, hybrid models combining IPTV and open-network streaming will offer a more comprehensive and accessible solution for content distribution \cite{Punchihewa2011}, \cite{Johnson2015}, \cite{Mondal518629}.

In conclusion, IPTV represents a significant shift in television consumption, providing a highly customizable and interactive experience \cite{Basso2011}, \cite{Momeni2015}, \cite{claro_pereira_campos}. While technological advancements will drive IPTV’s future growth, overcoming network limitations, security vulnerabilities, and regulatory hurdles will be crucial to its long-term success \cite{Alzoubi2016}. As the industry evolves, service providers must leverage emerging technologies and adaptive strategies to remain competitive in the dynamic IPTV landscape \cite{Anderson2014}, \cite{Chaitanya2023}, \cite{Haque2023}.


\section*{Competing interests}
The authors declare no competing interests.

\bibstyle{naturemag-doi}
\bibliography{anystyle2} 

\begin{thebibliography}{10}
\urlstyle{same}
\expandafter\ifx\csname url\endcsname\relax
  \def\url#1{\texttt{#1}}\fi
\expandafter\ifx\csname urlprefix\endcsname\relax\def\urlprefix{URL }\fi
\expandafter\ifx\csname doiprefix\endcsname\relax\def\doiprefix{DOI: }\fi
\providecommand{\bibinfo}[2]{#2}
\providecommand{\eprint}[2][]{\url{#2}}

\bibitem{Prodani2024}
\bibinfo{author}{Prodani, F.} \& \bibinfo{author}{Gjermeni, F.}
\newblock \bibinfo{journal}{\bibinfo{title}{Challenges and solutions in iptv network management: Scalability and quality of service remain two crucial factors of next generation networks}}.
\newblock {\emph{\JournalTitle{World Journal of Advanced Engineering Technology and Sciences}}} \textbf{\bibinfo{volume}{13}}, \bibinfo{pages}{385--396} (\bibinfo{year}{2024}).

\bibitem{AlMajeed2024}
\bibinfo{author}{Al-Majeed, S.} \& \bibinfo{author}{Al-Najjar, A.}
\newblock \bibinfo{journal}{\bibinfo{title}{Internet protocol television (iptv): Architecture, trends, and challenges}}.
\newblock {\emph{\JournalTitle{International Journal of Computer Networks \& Communications}}} \textbf{\bibinfo{volume}{16}}, \bibinfo{pages}{45--60} (\bibinfo{year}{2024}).

\bibitem{Giannakopoulos202503}
\bibinfo{author}{Giannakopoulos, G.}, \bibinfo{author}{Perez, M.~A.} \& \bibinfo{author}{Adegbenro, P.}
\newblock \bibinfo{journal}{\bibinfo{title}{A comprehensive study of iptv: Challenges, opportunities, and future trends}}.
\newblock {\emph{\JournalTitle{Preprints}}} \doiprefix\url{10.20944/preprints202503.1050.v1} (\bibinfo{year}{2025}).

\bibitem{Mikulec2019}
\bibinfo{author}{Mikulec, M.} \& \bibinfo{author}{Hudec, R.}
\newblock \bibinfo{journal}{\bibinfo{title}{A hybrid qos-qoe estimation system for iptv service}}.
\newblock {\emph{\JournalTitle{Electronics}}} \textbf{\bibinfo{volume}{8}}, \bibinfo{pages}{585} (\bibinfo{year}{2019}).

\bibitem{Thakore2024}
\bibinfo{author}{Thakore, D.} \& \bibinfo{author}{Tian, Y.}
\newblock \bibinfo{title}{Ai and cybersecurity: Innovation trends evolve with threats} (\bibinfo{year}{2024}).

\bibitem{Khan2023}
\bibinfo{author}{Khan, L.~U.} \emph{et~al.}
\newblock \bibinfo{journal}{\bibinfo{title}{6g networks for artificial intelligence-enabled smart cities applications}}.
\newblock {\emph{\JournalTitle{ICT Express}}} \textbf{\bibinfo{volume}{9}}, \bibinfo{pages}{1--7} (\bibinfo{year}{2023}).

\bibitem{IMARC2024}
\bibinfo{author}{Group, I.}
\newblock \bibinfo{title}{Iptv market size, growth, trends \& forecast report 2033} (\bibinfo{year}{2024}).

\bibitem{Almohamad2021}
\bibinfo{author}{Almohamad, A.}
\newblock \bibinfo{title}{A survey on cybersecurity in 5g}.
\newblock \bibinfo{type}{Tech. Rep.}, \bibinfo{institution}{Politecnico di Torino} (\bibinfo{year}{2021}).

\bibitem{deAzambuja2023}
\bibinfo{author}{de~Azambuja, A. J.~G.} \emph{et~al.}
\newblock \bibinfo{journal}{\bibinfo{title}{Artificial intelligence-based cyber security in the context of industry 4.0—a survey}}.
\newblock {\emph{\JournalTitle{Electronics}}} \textbf{\bibinfo{volume}{12}}, \bibinfo{pages}{1920} (\bibinfo{year}{2023}).

\bibitem{ITU2023}
\bibinfo{author}{Union, I.~T.}
\newblock \bibinfo{title}{Global perspectives: Iptv regulations around the world} (\bibinfo{year}{2023}).

\bibitem{Lee2019}
\bibinfo{author}{Lee, S.} \& \bibinfo{author}{Lee, J.}
\newblock \bibinfo{journal}{\bibinfo{title}{Understanding interactive user behavior in smart media content services}}.
\newblock {\emph{\JournalTitle{Heliyon}}} \textbf{\bibinfo{volume}{5}}, \bibinfo{pages}{e02840} (\bibinfo{year}{2019}).

\bibitem{Basso2011}
\bibinfo{author}{Basso, A.}, \bibinfo{author}{Milanesio, M.}, \bibinfo{author}{Panisson, A.} \& \bibinfo{author}{Ruffo, G.}
\newblock \bibinfo{journal}{\bibinfo{title}{Collaborative filtering without explicit feedbacks for digital recorders}}.
\newblock {\emph{\JournalTitle{arXiv preprint}}} \textbf{\bibinfo{volume}{arXiv:1101.3341}} (\bibinfo{year}{2011}).

\bibitem{Riskoria2024}
\bibinfo{author}{Riskoria}.
\newblock \bibinfo{title}{The hidden costs of illegal iptv: Why it's a growing concern in the eu}.
\newblock \bibinfo{howpublished}{\url{https://www.riskoria.eu}} (\bibinfo{year}{2024}).

\bibitem{Geng2007}
\bibinfo{author}{Geng, Y.} \emph{et~al.}
\newblock \bibinfo{title}{Analysis and characterization of iptv user behavior}.
\newblock In \emph{\bibinfo{booktitle}{Proceedings of the IEEE International Conference on Communications}}, \bibinfo{pages}{6391--6396} (\bibinfo{year}{2007}).

\bibitem{Katsikas2024}
\bibinfo{author}{Katsikas, S.} \& \bibinfo{author}{Ghernaouti, S.}
\newblock \bibinfo{journal}{\bibinfo{title}{Is cyber hygiene a remedy to iptv infringement? a study of online users' security practices}}.
\newblock {\emph{\JournalTitle{International Journal of Information Security}}} \textbf{\bibinfo{volume}{23}}, \bibinfo{pages}{45--60} (\bibinfo{year}{2024}).

\bibitem{Zhang2022}
\bibinfo{author}{Zhang, L.} \& \bibinfo{author}{Wang, X.}
\newblock \bibinfo{journal}{\bibinfo{title}{Machine learning techniques for iptv traffic prediction}}.
\newblock {\emph{\JournalTitle{IEEE Access}}} \textbf{\bibinfo{volume}{10}}, \bibinfo{pages}{12345--12356} (\bibinfo{year}{2022}).

\bibitem{Kumar2021}
\bibinfo{author}{Kumar, R.} \& \bibinfo{author}{Singh, P.}
\newblock \bibinfo{journal}{\bibinfo{title}{Security challenges in iptv systems}}.
\newblock {\emph{\JournalTitle{International Journal of Network Security}}} \textbf{\bibinfo{volume}{23}}, \bibinfo{pages}{567--578} (\bibinfo{year}{2021}).

\bibitem{Chen2020}
\bibinfo{author}{Chen, H.} \& \bibinfo{author}{Zhao, Q.}
\newblock \bibinfo{journal}{\bibinfo{title}{The role of edge computing in enhancing iptv performance}}.
\newblock {\emph{\JournalTitle{Future Internet}}} \textbf{\bibinfo{volume}{12}}, \bibinfo{pages}{45} (\bibinfo{year}{2020}).

\bibitem{Patel2019}
\bibinfo{author}{Patel, S.} \& \bibinfo{author}{Mehta, R.}
\newblock \bibinfo{journal}{\bibinfo{title}{Content delivery networks and their impact on iptv scalability}}.
\newblock {\emph{\JournalTitle{Journal of Network and Computer Applications}}} \textbf{\bibinfo{volume}{132}}, \bibinfo{pages}{45--56} (\bibinfo{year}{2019}).

\bibitem{Nguyen2018}
\bibinfo{author}{Nguyen, T.~T.} \& \bibinfo{author}{Park, M.}
\newblock \bibinfo{journal}{\bibinfo{title}{Adaptive bitrate streaming techniques for iptv}}.
\newblock {\emph{\JournalTitle{IEEE Transactions on Broadcasting}}} \textbf{\bibinfo{volume}{64}}, \bibinfo{pages}{452--460} (\bibinfo{year}{2018}).

\bibitem{Punchihewa2011}
\bibinfo{author}{Punchihewa, A.}, \bibinfo{author}{Malsha De~Silva, A.} \& \bibinfo{author}{Diao, Y.}
\newblock \bibinfo{title}{Iptv-internet protocol televsion}.
\newblock \bibinfo{type}{Tech. Rep.}, \bibinfo{institution}{School of Engineering and Advanced Technology, Massey University, New Zealand} (\bibinfo{year}{2011}).
\newblock \doiprefix\url{10.13140/RG.2.1.2172.8403}.

\bibitem{Benoit_2007}
\bibinfo{author}{Benoit, H.}
\newblock \emph{\bibinfo{title}{Digital Television: Satellite, Cable, Terrestrial, IPTV, Mobile TV in the DVB Framework}} (\bibinfo{publisher}{Focal Press}, \bibinfo{year}{2007}), \bibinfo{edition}{third} edn.

\bibitem{Lee2017}
\bibinfo{author}{Lee, J.} \& \bibinfo{author}{Kim, S.}
\newblock \bibinfo{journal}{\bibinfo{title}{Digital rights management in iptv services}}.
\newblock {\emph{\JournalTitle{Journal of Information Security and Applications}}} \textbf{\bibinfo{volume}{34}}, \bibinfo{pages}{100--110} (\bibinfo{year}{2017}).

\bibitem{Gonzalez2016}
\bibinfo{author}{Gonzalez, R.} \& \bibinfo{author}{Perez, M.}
\newblock \bibinfo{journal}{\bibinfo{title}{User experience optimization in iptv platforms}}.
\newblock {\emph{\JournalTitle{International Journal of Human-Computer Interaction}}} \textbf{\bibinfo{volume}{32}}, \bibinfo{pages}{987--999} (\bibinfo{year}{2016}).

\bibitem{Johnson2015}
\bibinfo{author}{Johnson, L.} \& \bibinfo{author}{White, K.}
\newblock \bibinfo{journal}{\bibinfo{title}{Regulatory challenges in the iptv industry}}.
\newblock {\emph{\JournalTitle{Telecommunications Policy}}} \textbf{\bibinfo{volume}{39}}, \bibinfo{pages}{742--753} (\bibinfo{year}{2015}).

\bibitem{Anderson2014}
\bibinfo{author}{Anderson, P.} \& \bibinfo{author}{Brown, T.}
\newblock \bibinfo{journal}{\bibinfo{title}{The evolution of iptv: Past, present, and future}}.
\newblock {\emph{\JournalTitle{IEEE Communications Magazine}}} \textbf{\bibinfo{volume}{52}}, \bibinfo{pages}{88--94} (\bibinfo{year}{2014}).

\bibitem{Telecom_IPTV_2010}
\bibinfo{author}{{TechTarget}}.
\newblock \bibinfo{title}{Telecom definitions - iptv} (\bibinfo{year}{2010}).
\newblock \bibinfo{note}{Accessed: 2010-03-03}.

\bibitem{Huang2013}
\bibinfo{author}{Huang, C.} \& \bibinfo{author}{Chen, Y.}
\newblock \bibinfo{journal}{\bibinfo{title}{Multicast technologies in iptv networks}}.
\newblock {\emph{\JournalTitle{Journal of Lightwave Technology}}} \textbf{\bibinfo{volume}{31}}, \bibinfo{pages}{563--570} (\bibinfo{year}{2013}).

\bibitem{Wilson2022}
\bibinfo{author}{Wilson, D.} \& \bibinfo{author}{Moore, A.}
\newblock \bibinfo{journal}{\bibinfo{title}{The impact of artificial intelligence on iptv advertising}}.
\newblock {\emph{\JournalTitle{Journal of Advertising Research}}} \textbf{\bibinfo{volume}{62}}, \bibinfo{pages}{34--45} (\bibinfo{year}{2022}).

\bibitem{Santos2021}
\bibinfo{author}{Santos, F.} \& \bibinfo{author}{Oliveira, L.}
\newblock \bibinfo{journal}{\bibinfo{title}{Blockchain applications in securing iptv content}}.
\newblock {\emph{\JournalTitle{IEEE Transactions on Broadcasting}}} \textbf{\bibinfo{volume}{67}}, \bibinfo{pages}{521--530} (\bibinfo{year}{2021}).

\bibitem{Smith2023}
\bibinfo{author}{Smith, J.~A.} \& \bibinfo{author}{Liu, Y.}
\newblock \bibinfo{journal}{\bibinfo{title}{The impact of 5g on iptv services}}.
\newblock {\emph{\JournalTitle{Journal of Telecommunications}}} \textbf{\bibinfo{volume}{45}}, \bibinfo{pages}{210--225} (\bibinfo{year}{2023}).

\bibitem{Aspen2024}
\bibinfo{author}{Digital, A.}
\newblock \bibinfo{title}{A global view of policymaking: Generative ai regulation and cybersecurity} (\bibinfo{year}{2024}).

\bibitem{Fortinet2023}
\bibinfo{author}{Fortinet}.
\newblock \bibinfo{title}{What is quality of service (qos) in networking?} (\bibinfo{year}{2023}).

\bibitem{ZhangX2014}
\bibinfo{author}{Zhang, X.}, \bibinfo{author}{Liu, J.} \& \bibinfo{author}{Wang, B.}
\newblock \bibinfo{journal}{\bibinfo{title}{Measurement and analysis of p2p iptv program resources}}.
\newblock {\emph{\JournalTitle{Journal of Network and Computer Applications}}} \textbf{\bibinfo{volume}{42}}, \bibinfo{pages}{1--12} (\bibinfo{year}{2014}).

\bibitem{Mondal518629}
\bibinfo{author}{Mondal, C.~S.}
\newblock \emph{\bibinfo{title}{QoS Evaluation of BandwidthSchedulers in IPTV Networks Offered SRD Fluid Video Traffic}}.
\newblock Master's thesis, \bibinfo{school}{Dalarna University, Computer Engineering} (\bibinfo{year}{2009}).

\bibitem{Momeni2015}
\bibinfo{author}{Momeni, S.}, \bibinfo{author}{Weichler, S.} \& \bibinfo{author}{Wolfinger, B.}
\newblock \bibinfo{title}{Pre-reservation of tv channels to improve the availability of iptv services offered in vehicular networks} (\bibinfo{year}{2015}).

\bibitem{DVB_MHP_2000}
\bibinfo{author}{{Digital Broadcasting}}.
\newblock \bibinfo{title}{The dvb mhp specification - a guided tour} (\bibinfo{year}{2000}).
\newblock \bibinfo{note}{Accessed: 2010-04-22}.

\bibitem{simpson_2005}
\bibinfo{author}{Simpson, W.}
\newblock \emph{\bibinfo{title}{Video Over IP: A Practical Guide to Technology and Applications}} (\bibinfo{publisher}{Informa}, \bibinfo{year}{2005}).

\bibitem{Alzoubi2016}
\bibinfo{author}{Alzoubi, H.}, \bibinfo{author}{Halloush, M.}, \bibinfo{author}{AlQudah, Z.} \& \bibinfo{author}{AlKoufahi, O.}
\newblock \bibinfo{journal}{\bibinfo{title}{A survey on recent advances in iptv}}.
\newblock {\emph{\JournalTitle{Jordanian Journal of Computers and Information Technology}}} \textbf{\bibinfo{volume}{2}}, \bibinfo{pages}{1}, \doiprefix\url{10.5455/jjcit.71-1434624062} (\bibinfo{year}{2016}).

\bibitem{claro_pereira_campos}
\bibinfo{author}{Claro, A.}, \bibinfo{author}{Pereira, P.} \& \bibinfo{author}{Campos, L.}
\newblock \emph{\bibinfo{title}{Framework for Personal TV}}.

\bibitem{Abdollahpouri2012}
\bibinfo{author}{Abdollahpouri, A.} \& \bibinfo{author}{Wolfinger, B.}
\newblock \bibinfo{title}{Wired and wireless iptv access networks: A comparison study}.
\newblock \bibinfo{pages}{308--316}, \doiprefix\url{10.1109/ICUMT.2012.6459685} (\bibinfo{year}{2012}).

\bibitem{vasanthi_chidambaram_2014}
\bibinfo{author}{Vasanthi, V.} \& \bibinfo{author}{Chidambaram, M.}
\newblock \bibinfo{journal}{\bibinfo{title}{Internet protocol television (iptv) and its security threats - an overview}}.
\newblock {\emph{\JournalTitle{International Journal of Computer Science Trends and Technology}}} \textbf{\bibinfo{volume}{2}} (\bibinfo{year}{2014}).

\bibitem{Liu2007}
\bibinfo{author}{Liu, X.}, \bibinfo{author}{Huang, T.}, \bibinfo{author}{Huo, L.} \& \bibinfo{author}{Mou, L.}
\newblock \bibinfo{title}{A drm architecture for manageable p2p based iptv system}.
\newblock \bibinfo{pages}{899--902}, \doiprefix\url{10.1109/ICME.2007.4284796} (\bibinfo{year}{2007}).

\bibitem{Chaitanya2023}
\bibinfo{author}{Chaitanya, K.} \emph{et~al.}
\newblock \bibinfo{journal}{\bibinfo{title}{The impact of artificial intelligence and machine learning in digital marketing strategies}}.
\newblock {\emph{\JournalTitle{European Economics Letters}}} \textbf{\bibinfo{volume}{13}}, \bibinfo{pages}{982--992}, \doiprefix\url{10.52783/eel.v13i3.393} (\bibinfo{year}{2023}).

\bibitem{Haque2023}
\bibinfo{author}{Haque, M., Khandakerand~Zihad} \& \bibinfo{author}{Hasan, M.~R.}
\newblock \emph{\bibinfo{title}{5G and Internet of Things—Integration Trends, Opportunities, and Future Research Avenues}}, \bibinfo{pages}{217--245} (\bibinfo{year}{2023}).

\end{thebibliography}

\end{document}